\documentclass[10pt,journal]{IEEEtran}
\usepackage{lineno,hyperref}
\usepackage{amssymb}
\usepackage{mathtools, cuted}
\usepackage{color, comment, verbatim}
\usepackage{latexsym}
\usepackage{enumerate}
\usepackage{caption}
\usepackage{subcaption}
\usepackage{algorithm, algpseudocode}
\usepackage[dvipsnames]{xcolor}
\ifCLASSINFOpdf
\else
\fi
\begin{document}

\title{Model Selection in High-Dimensional Block-Sparse Linear Regression}

\author{Prakash~B. Gohain,~\IEEEmembership{Student Member,~IEEE,}
        Magnus~Jansson,~\IEEEmembership{Senior Member,~IEEE.}
        % <-this % stops a space
\thanks{This research was supported in part by the European Research Council (ERC) under the European Union’s Horizon 2020 research and innovation programme, grant agreement No.  742648. 

The authors are with the Division of Information Science and Engineering, KTH Royal Institute of Technology, Stockholm SE-10044, Sweden
e-mail: pbg@kth.se, janssonm@kth.se.}% <-this % stops a space
}


\markboth{Journal of \LaTeX\ Class Files,~Vol.~14, No.~8, August~2022}%
{Shell \MakeLowercase{\textit{et al.}}: Bare Demo of IEEEtran.cls for IEEE Journals}



\maketitle

\begin{abstract}
Model selection is an indispensable part of data analysis dealing very frequently with fitting and prediction purposes.
In this paper, we tackle the problem of model selection in a general linear regression where the parameter matrix possesses a block-sparse structure, i.e., the non-zero entries occur in clusters or blocks and the number of such non-zero blocks is very small compared to the parameter dimension.
Furthermore, a high-dimensional setting is considered where the parameter dimension is quite large compared to the number of available measurements. To perform model selection in this setting, we present an information criterion that is a generalization of the Extended Bayesian Information Criterion-Robust (EBIC-R) and it takes into account both the block structure and the high-dimensionality scenario. The analytical steps for deriving the EBIC-R for this setting are provided. Simulation results show that the proposed method performs considerably better than the existing state-of-the-art methods and achieves empirical consistency at large sample sizes and/or at high-SNR. 
\end{abstract}


% \fi
\IEEEpeerreviewmaketitle

%----------------------%
\section{Introduction}
%----------------------%
%Selecting the best model/subset in the high-dimensional linear regression model has been an active research topic for a long time now. With the advent of big data, encountering high-dimensional datasets is a common phenomenon in many fields of science, engineering, finance, marketing, biology, etc. This provided a massive impulse to explore efficient tools and techniques, particularly in the field of data analysis. In this regard, model selection plays a central role in data analysis, and as such developing efficient and robust methods for model selection is crucial for reliable data analysis \cite{MOS_overview_2018}. 
Selecting the best model/subset in the high-dimensional (HD) linear regression has been an active research topic for a long time now. In this context, methods based on Information Criterion (IC) have played a pivotal role ever since Akaike proposed the famous Akaike IC \cite{AIC_Akaike1974}. Since then, IC-based model selection has come a long way with the frequent appearance of more developed and sophisticated methods \cite{MS_Rao_2001, MOS_review_Stoica2004, MS_review_chakrbarti_2011}. However, classical methods do not fare well in HD scenarios and quite often lead to overfitted models with false parameters.
In the present era, popular IC-based methods for model selection in the HD setting include extended Bayesian IC (EBIC) \cite{EBIC2008}, extended Fisher IC (EFIC) \cite{EFIC2018}, and extended BIC-Robust (EBIC$_\text{R}$) \cite{Gohain2022EBICR, gohain2022_EBICR_IEEEtrans}. 

Apart from the methods mentioned above, there are other non-IC based model selection approaches in the HD regime such as the Residual-Ratio-Thresholding (RRT) \cite{RRT-2018} and the Multi-Beta-Test (MBT) \cite{gohain2020_MBT}, both of which are based on the hypothesis testing framework used along with a greedy variable selection method such as orthogonal matching pursuit (OMP) \cite{OMP_Cai2011}. Some recent methods also include significance test of the LASSO \cite{significance-test-lasso} and knock-off-filters \cite{knock-off-filters}. Another popular method is cross-validation (CV) \cite{CV_shao1993, CV_picard1984}. However, CV-based procedures can be computationally intensive, and their performance in the HD setting is not satisfactory \cite{CV_p_greater_N, CV-drawback, CV-survey}.

In this paper, we consider model selection in a general linear regression where the nonzero coefficients in the parameter matrix occur in clusters (or groups). Such signals are referred to as block-sparse \cite{eldar2009block, eldar2009robust, B-OMP-Eldar}.
Block-sparsity inherently arises in a variety of scenarios. For example in multi-band signals
\cite{multiband-shali2009, multiband-mishali2010theory}, in the recovery of signals from compressed microarray measurements
\cite{gene-expression-parvaresh2008recovering}, in the multiple measurement vector (MMV) problem 
\cite{MMV-cotter2005sparse, MMV-chen2006theoretical, MMV-Arash, MMV-eldar2009average}. Furthermore, as shown in \cite{eldar2009block} and \cite{eldar2009robust}, the block-sparsity model can be used to handle the issue of sampling signals that lie in a union of subspaces
%\cite{subspaces-casazza2004frames, subspaces-lu2008sampling, subspaces-blumensath2009sampling, subspaces-eldar2009compressed,}
\cite{subspaces-mishali2009blind, subspaces-gedalyahu2010time}.

A recent method for model selection in block-sparse HD linear regression is the Generalized RRT (GRRT) \cite{kallummil2022GRRT}. GRRT is an extension of RRT \cite{RRT-2018} developed to treat the block-sparse structure in linear regression. The authors also present a new approach that allows GRRT to perform model selection in non-monotonic predictor sequences generated by LASSO \cite{lasso_tibshirani1996}.
However, the IC-based methods are not designed to take into account the block structure during model selection. Hence, in their current form, they cannot be applied directly without tailoring them to incorporate the block nature of the underlying linear model into the criterion. In this paper, the main goal is to develop an IC-based model selection method for the general linear regression model assuming a block-sparse structure and a HD setting.

In the paper, matrices and vectors are denoted by boldface letters. The notation $(\cdot)^T$ stands for transpose.  $\mathbf{I}_N$ is an $N\times N$ identity matrix. $\mathbf{\Pi}(\mathbf{A}) = \mathbf{A} (\mathbf{A}^T\mathbf{A})^{-1}\mathbf{A}^T$ represents the orthogonal projection matrix on the column space of $\mathbf{A}$ and $\mathbf{\Pi}^\perp(\mathbf{A}) = \mathbf{I}_N - \mathbf{\Pi}(\mathbf{A})$ the orthogonal projection matrix on the null space of $\mathbf{A}^T$. The notation $\big\lvert \mathbf{X} \big\rvert$ denotes the determinant of the matrix $\mathbf{X}$, $\lVert \cdot \rVert_2$ denotes the Euclidean norm and $\lVert \cdot \rVert_F$ the Frobenius norm. $X\sim \mathcal{N}(\mu,\sigma^2)$ signifies a Gaussian distributed random variable with mean $\mu$ and variance $\sigma^2$. The symbol $\otimes$ represents the Kronecker product and vec$(\mathbf{A})$ signifies the vectorization of the matrix $\mathbf{\mathbf{A}}$.

%The rest of the paper is organized as follows: In Section II the problem statement is presented that shows the considered model and the model selection problem at hand.  In Section III we present the proposed method and provide the necessary derivation steps to arrive at the method. Section IV shows the simulation results and Section V concludes the paper.
%--------------------------%
\section{Problem Statement}
%--------------------------%
Technically there can be four different linear regression structures depending on the configuration of the parameter matrix (or vector). They are (a) single measurement vector (SMV), (b) block single measurement vector (BSMV), (c) multiple measurement vector (MMV) and (d) block multiple measurement vector (BMMV).
For example, as mentioned in \cite{kallummil2022GRRT}, SMV models are used in wireless signal detection \cite{smv-grrt-choi2017detection}, MMV models in Electroencephalogram (EEG) \cite{MMV-grrt-2007compressed}, BSMV models in multi pitch estimation \cite{BSMV-grrt-kronvall2017group} and BMMV models in face recognition \cite{BMMV-grrt-fedorov2016robust}.
Here, we consider the BMMV model, since it is the general setting and the rest of the models are special cases of BMMV. The BMMV model is as follows:
\begin{equation}
    \mathbf{Y} = \mathbf{AX} + \mathbf{W},\label{eq:block-linear-model}
\end{equation}
where, $\mathbf{Y} \in \mathbb{R}^{N\times L}$ is the observed response matrix, $\mathbf{A} \in \mathbb{R}^{N\times p}$ is the design matrix, $\mathbf{X} \in \mathbb{R}^{p\times L}$ is the unknown parameter matrix and $\mathbf{W} \in \mathbb{R}^{N \times L}$ is the noise/error matrix, whose elements are assumed to be i.i.d. and $\mathbf{W}[i,j]\sim \mathcal{N}(0,\sigma^2)$. 
The $p$ rows of $\mathbf{X}$ are divided into $p_B = p/L_B$ unique blocks of equal size $L_B$. Each of these $p_B$ blocks of size $L_B \times L$ is non-zero or zero at once. The block size $L_B$ is assumed to be known a-\textit{priori}. The $j$th block consists of the rows of $\mathbf{X}$ indexed by $\mathcal{I}_j = \{(j-1)L_B+1, (j-1)L_B+2,\ldots,jL_B \}$. We denote the true block support of $\mathbf{X}$ as $\mathcal{S}_B= \left\{ j\in [p_B]:\mathbf{X}[\mathcal{I}_j,:]\neq \mathbf{0}_{L_B\times L} \right\}$. 
Also, $\mathbf{X}$ is assumed to be block-sparse such that $K_B = card(\mathcal{S}_B)\ll p_B$. Table \ref{tab:types-LR-model} shows the different linear regression structures. 
The goal of model selection herein is estimating $\mathcal{S}_B$ given $\mathbf{Y}$ and $\mathbf{A}$. 

\begin{table}[t]
\begin{center}    
\begin{tabular}{|c|c|c|}
    \hline
    Type    &   Specifications & $dim(\mathbf{Y})$, $dim(\mathbf{X})$  \\
    \hline 
     SMV    & $L=1$, $L_B =1$, $p_B = p$  & $N\times 1$, $p\times 1$   \\
     MMV    & $L>1$, $L_B =1$, $p_B = p$  &  $N\times L$, $p\times L$  \\
     BSMV   &  $L=1$, $L_B >1$, $p_B = p/L_B$  & $N\times 1$, $p\times 1$  \\
     BMMV   &  $L>1$, $L_B >1$, $p_B = p/L_B$  & $N\times L$, $p\times L$  \\
     \hline     
\end{tabular}
\end{center}
    \caption{Types of linear regression structures.}
    \label{tab:types-LR-model}
\end{table}

The model selection procedure can be divided into two stages: (i) In the first stage, we pick a competent set of candidate models using an appropriate predictor/subset selection algorithm up to maximum cardinality $K$ under the assumption that $K_B \leq K \ll N$. (ii) In the second stage, we estimate the true model using a suitable model selection criterion.
Let us denote $\mathcal{I}_B$ as the block support of a candidate model such that $card(\mathcal{I}_B) = k_B$, where $k_B\in \{1,2,\ldots,p_B\}$. Then we can reformulate the linear model in (\ref{eq:block-linear-model}) as
\begin{equation}
\mathcal{H}_{\mathcal{I}_B} : \mathbf{Y} = \mathbf{A}_{\mathcal{I}_B}\mathbf{X}_{\mathcal{I}_B} + \mathbf{W}_{\mathcal{I}_B}, \label{eq:block-linear-model-hypothesis} 
\end{equation}
where $\mathcal{H}_{\mathcal{I}_B}$ signifies the hypothesis that the data $\mathbf{Y}$ is actually produced in accordance with (\ref{eq:block-linear-model-hypothesis}), $\mathbf{A}_{\mathcal{I}_B}\in \mathbb{R}^{N\times (k_B L_B)}$ is the sub-matrix consisting of columns from the known matrix $\mathbf{A}$ with block support $\mathcal{I}_B \subseteq \{1,2,\ldots,p_B\}$, $\mathbf{X}_{\mathcal{I}_B} \in \mathbb{R}^{(k_BL_B)\times L}$ is the corresponding unknown parameter coefficient matrix, and $\mathbf{W}_{\mathcal{I}_B} \in \mathbb{R}^{N\times L}$ is the associated noise matrix.

%------------------------%
\section{Proposed Method}
%------------------------%
In this section, we provide the necessary steps to derive the EBIC$_\text{R}$ to perform model selection in block-sparse general linear regression for the BMMV scenario. The further analysis assumes the following property of the design matrix $\mathbf{A}$  \cite{Stoica_BIC_SNR, Schmidt_2011_Consistency_of_MDL, GOHAIN-BIC-R}
\begin{equation}
    \lim_{N\to \infty} \left\{ N^{-1}(\mathbf{A}^T_{\mathcal{I}_B}\mathbf{A}_{\mathcal{I}_B})\right\} =\mathbf{M}_{\mathcal{I}_B}=\mathcal{O}(1),
    \label{eq:positive_definite}
\end{equation}
where $\mathbf{M}_{\mathcal{I}_B}$ is a $(k_BL_B \times k_BL_B)$ positive definite matrix and bounded as $N \to \infty$.
The assumption in (\ref{eq:positive_definite}) holds true in many cases but not all (see \cite{djuric1998asymptotic, Stoica_BIC_SNR} for more details). 

To arrive at the EBIC$_\text{R}$ for the BMMV model, we first reformulate the linear model in (\ref{eq:block-linear-model-hypothesis}) into vector form as
\begin{equation}
  \text{vec}(\mathbf{Y}) = \mathbf{I}_L\otimes \mathbf{A}_{\mathcal{I}_B} \text{vec}(\mathbf{X}_{\mathcal{I}_B}) + \text{vec}(\mathbf{W}_{\mathcal{I}_B}). \label{eq:vector-system-model-1}
\end{equation}
This step allows us to utilize the same derivation steps as in $\cite{Gohain2022EBICR}$ 
without the need to carry out the analysis from scratch. Also, (\ref{eq:vector-system-model-1}) is technically equivalent to (\ref{eq:block-linear-model-hypothesis}), hence we do not alter the underlying original linear model but just restructure it for our convenience.
Now, let $\mathbf{y}=\text{vec}(\mathbf{Y}) \in \mathbb{R}^{NL\times 1}$, $\breve{\mathbf{A}}_\mathcal{I} = \mathbf{I}_L\otimes \mathbf{A}_{\mathcal{I}_B}\in \mathbb{R}^{NL \times k_BL_BL}$, $\mathbf{x}_\mathcal{I} = \text{vec}(\mathbf{X}_{\mathcal{I}_B})\in \mathbb{R}^{k_BL_BL\times 1}$ and $\mathbf{e}_\mathcal{I} = \text{vec}(\mathbf{W}_{\mathcal{I}_B}) \in \mathbb{R}^{NL \times 1}$. The elements of $\mathbf{e}_\mathcal{I}$ are i.i.d. and $\mathbf{e}_\mathcal{I} \sim \mathcal{N}(\mathbf{0},\sigma^2_\mathcal{I}\mathbf{I}_{NL})$.
Then, we can rewrite (\ref{eq:vector-system-model-1}) as
\begin{equation}
\mathcal{H}_{{\mathcal{I}}} : \mathbf{y} = \breve{\mathbf{A}}_{{\mathcal{I}}}\mathbf{x}_{{\mathcal{I}}} + \mathbf{e}_{{\mathcal{I}}}, \label{eq:LR_model_2}
\end{equation}
where $\mathcal{I}\subseteq \{1,2,\ldots,pL\}$.
Then the pdf of $\mathbf{y}$ under $\mathcal{H}_\mathcal{I}$ is
\begin{equation}
    p\left(\mathbf{y}\lvert \boldsymbol{\theta}_\mathcal{I},\mathcal{H}_\mathcal{I}\right) = \frac{ \exp\{- \lVert \mathbf{y} -{\breve{\mathbf{A}}}_\mathcal{I}\mathbf{x}_\mathcal{I} \rVert^2_2/2\sigma^2_\mathcal{I}\} }{(2\pi \sigma^2_\mathcal{I})^{NL/2}}, \label{eq:pdf_y}   
\end{equation}
where $\boldsymbol{\theta}_\mathcal{I} = [\mathbf{x}^T_\mathcal{I}, \sigma^2_\mathcal{I}]^T$ is the vector of all the unknown parameters of the model under $\mathcal{H}_\mathcal{I}$. The maximum likelihood estimates (MLE) $\boldsymbol{\hat{\theta}}_\mathcal{I} = [\hat{\mathbf{x}}^T_\mathcal{I}, \hat{\sigma}^2_\mathcal{I}]^T$
are obtained as \cite{S_Kay_estimation_book}
\begin{equation}
    \hat{\mathbf{x}}_\mathcal{I} = \left({\breve{\mathbf{A}}}^T_\mathcal{I}{\breve{\mathbf{A}}}_\mathcal{I}\right)^{-1}{\breve{\mathbf{A}}}^T_\mathcal{I} \mathbf{y} \quad \& \quad  \hat{\sigma}^2_\mathcal{I}=\frac{\mathbf{y}^T\mathbf{\Pi}^\perp(\breve{\mathbf{A}}_\mathcal{I})\mathbf{y}}{NL}.\label{eq:MLEs_theta}
\end{equation}
EBIC$_\text{R}$ is derived under the Bayesian framework of model selection, which starts with deriving the maximum a-posteriori (MAP) criterion and ending with the final EBIC$_\text{R}$ after suitable modifications and reasonable assumptions. We follow similar steps as in \cite{Gohain2022EBICR, gohain2022_EBICR_IEEEtrans}, but incorporate the multi-measurement and block structure into it.
Let us denote the prior pdf of the parameter vector $\boldsymbol{\theta}_\mathcal{I}$ as $p(\boldsymbol{\theta}_\mathcal{I}|\mathcal{H}_\mathcal{I})$, the marginal of $\mathbf{y}$ as $p(\mathbf{y}|\mathcal{H}_\mathcal{I})$ and the prior probability of the model with support 
$\mathcal{I}$ as $\Pr(\mathcal{H}_\mathcal{I})$.
% Let $p(\boldsymbol{\theta}_\mathcal{I}|\mathcal{H}_\mathcal{I})$ denote the prior pdf of the parameter vector $\boldsymbol{\theta}_\mathcal{I}$ under $\mathcal{H}_\mathcal{I}$. Then we have the joint probability
% \begin{equation}
%     p(\mathbf{y},\boldsymbol{\theta}_\mathcal{I}|\mathcal{H}_\mathcal{I}) = p(\mathbf{y}|\boldsymbol{\theta}_\mathcal{I},\mathcal{H}_\mathcal{I})p(\boldsymbol{\theta}_\mathcal{I}|\mathcal{H}_\mathcal{I}) \label{eq:joint_density}
% \end{equation}
% and the marginal distribution of $\mathbf{y}$ is
% \begin{equation}
%     p(\mathbf{y}|\mathcal{H}_\mathcal{I}) = \int p(\mathbf{y}|\boldsymbol{\theta}_\mathcal{I},\mathcal{H}_\mathcal{I})p(\boldsymbol{\theta}_\mathcal{I}|\mathcal{H}_\mathcal{I})d\boldsymbol{\theta}_\mathcal{I}. \label{eq:marginal}
% \end{equation}
% The posterior probability $\Pr(\mathcal{H}_\mathcal{I}|\mathbf{y})$ is given by
% \begin{equation}
%     \Pr(\mathcal{H}_\mathcal{I}|\mathbf{y}) = \frac{p(\mathbf{y}|\mathcal{H}_\mathcal{I})\Pr\left(\mathcal{H}_\mathcal{I}\right)}{p(\mathbf{y})},\label{eq:posterior_prob}
% \end{equation}
% where $\Pr(\mathcal{H}_\mathcal{I})$ is the prior probability of the model with support $\mathcal{I}$. The MAP estimator picks the model having the largest $\Pr(\mathcal{H}_\mathcal{I}|\mathbf{y})$.
% However, note that the $p(\mathbf{y})$ is a normalizing factor and independent of $\mathcal{I}$. 
Then the MAP estimate of the true support $\mathcal{S} \subseteq \{1,2,\ldots pL\}$ is equivalently given by \cite{Stoica_BIC_SNR, GOHAIN-BIC-R}
\begin{equation}
    \hat{\mathcal{S}}_\text{MAP} = \underset{\mathcal{I}}{\arg \max}\ \Big \{ \ln p(\mathbf{y}|\mathcal{H}_\mathcal{I}) + \ln \Pr\left(\mathcal{H}_\mathcal{I}\right) \Big \}.
\end{equation}
%Under the assumption that $N$ and/or SNR are large, an approximation of  $\ln p(\mathbf{y}|\mathcal{H}_\mathcal{I})$ is obtained using a second order Taylor series expansion, which gives (see \cite{Stoica_BIC_SNR, GOHAIN-BIC-R} for details)
Applying a second order Taylor series expansion, an approximation of $\ln p(\mathbf{y}|\mathcal{H}_\mathcal{I})$ is obtained under the presumption that $N$ is large or/and SNR is high (see \cite{Stoica_BIC_SNR, GOHAIN-BIC-R} for details)
\begin{equation}
    \begin{split}
    \ln p(\mathbf{y}|\mathcal{H}_\mathcal{I}) \approx  \ln p(\mathbf{y}|\hat{\boldsymbol{\theta}}_\mathcal{I},\mathcal{H}_\mathcal{I})+ \ln p(\hat{\boldsymbol{\theta}}_\mathcal{I}|\mathcal{H}_\mathcal{I})\\
    +\frac{k_BL_BL+1}{2}\ln(2\pi)-\frac{1}{2}\ln \big|\hat{\mathbf{F}}_\mathcal{I}\big|. \label{eq:approx_ln_pdf_of_y}
    \end{split}
\end{equation}
$\hat{\mathbf{F}}_\mathcal{I}$ is the sample Fisher information matrix \cite{S_Kay_estimation_book} under $\mathcal{H}_\mathcal{I}$ evaluated at the MLE, hence (see \cite{Stoica_BIC_SNR, GOHAIN-BIC-R}) 
%$\hat{\mathbf{F}}_\mathcal{I}$ is the sample Fisher information matrix under $\mathcal{H}_\mathcal{I}$ given as \cite{S_Kay_estimation_book}
% \begin{equation}
%     \hat{\mathbf{F}}_\mathcal{I} = -\frac{\partial^2\ln p(\mathbf{y}|{\boldsymbol{\theta}}_\mathcal{I},\mathcal{H}_\mathcal{I})}{\partial \boldsymbol{\theta}_\mathcal{I}\partial \boldsymbol{\theta}_\mathcal{I}^T}\bigg \lvert_{{\boldsymbol{\theta}}_\mathcal{I} = \hat{\boldsymbol{\theta}}_\mathcal{I}}.\label{eq:FIM_raw}
% \end{equation}
%Evaluating (\ref{eq:FIM_raw}) using (\ref{eq:pdf_y}) and (\ref{eq:MLEs_theta}) we get \cite{Stoica_BIC_SNR}
\begin{equation}
    \hat{\mathbf{F}}_\mathcal{I} = 
    \begin{bmatrix}
    \frac{1}{\hat{\sigma}^2_\mathcal{I}}{\breve{\mathbf{A}}}_\mathcal{I}^T{\breve{\mathbf{A}}}_\mathcal{I} & \mathbf{0} \\
    \mathbf{0} & \frac{NL}{2\hat{\sigma}^4_\mathcal{I}}
    \end{bmatrix}.\label{eq:FIM_matrix}
\end{equation}
From the linear model in \ref{eq:LR_model_2} we have
\begin{equation}
    -2\ln p(\mathbf{y}|\boldsymbol{\hat{\theta}}_\mathcal{I},\mathcal{H}_\mathcal{I}) = NL\ln \hat{\sigma}^2_\mathcal{I} +\text{const}.\label{eq:pdf_LR_MLE}
\end{equation}
Now, using (\ref{eq:pdf_LR_MLE}), it is possible to rewrite (\ref{eq:approx_ln_pdf_of_y}) as
\begin{equation}
    \begin{split}
     -2\ln p(\mathbf{y}|\mathcal{H}_\mathcal{I}) \approx NL\ln \hat{\sigma}^2_\mathcal{I} + \ln \big|\hat{\mathbf{F}}_\mathcal{I}\big | - 2\ln p(\hat{\boldsymbol{\theta}}_\mathcal{I}|\mathcal{H}_\mathcal{I})\\
     - k_BL_BL\ln 2\pi +\text{const.}
    \end{split}
\end{equation}
Furthermore, the prior term in (\ref{eq:approx_ln_pdf_of_y}), i.e., $\ln p(\hat{\boldsymbol{\theta}}_\mathcal{I}|\mathcal{H}_\mathcal{I})$, is ignored under the pretext that it is flat and uninformative.
Thus, discarding the constants and the terms not dependent on the block model dimension $k_B$, we can equivalently reformulate the MAP-based model estimate as
\begin{align}
     \hat{\mathcal{S}}_\text{MAP} = \underset{\mathcal{I}}{\arg \min}\Big\{NL\ln \hat{\sigma}^2_\mathcal{I} + \ln & \big| \hat{\mathbf{F}}_\mathcal{I}\big| - k_BL_BL\ln 2\pi\nonumber\\
     &-2\ln \Pr\left(\mathcal{H}_\mathcal{I}\right)  \Big \}.\label{eq:MAP_estimate_final}
\end{align}
EBIC$_\text{R}$ is derived from (\ref{eq:MAP_estimate_final}) with some further modifications and approximations. The two key terms that require further analysis are $\ln \lvert \hat{\mathbf{F}}_\mathcal{I} \rvert$ and the prior term $\Pr(\mathcal{H}_\mathcal{I}) $.
First, we perform normalization of $\hat{\mathbf{F}}_\mathcal{I}$ under both large-$N$ and high-SNR assumption. For this we factorize the $\ln \big\lvert \hat{\mathbf{F}}_\mathcal{I}\big\rvert$term in a similar manner as performed in \cite{Gohain2022EBICR, gohain2022_EBICR_IEEEtrans, GOHAIN-BIC-R}
\begin{align}
    \ln{\big|\hat{\mathbf{F}}_\mathcal{I}\big|} = & \ln \left[\big|\mathbf{Q}\big|\left \lvert \mathbf{Q}^{-1/2}\hat{\mathbf{F}}_\mathcal{I} \mathbf{Q}^{-1/2}\right \rvert\right ]\nonumber\\
    = & \ln|\mathbf{Q}| + \ln {\Big | \mathbf{Q}^{-1/2} \mathbf{\hat{F}}_\mathcal{I} \mathbf{Q}^{-1/2}\Big |}.\label{eq:ln_det_FIM_EBIC_R}
\end{align}
The objective here is to choose a suitable $\mathbf{Q}$ matrix that normalizes $\hat{\mathbf{F}}_\mathcal{I}$ such that the second term in (\ref{eq:ln_det_FIM_EBIC_R}) is $\mathcal{O}(1)$, i.e., it should be bounded as $N \to \infty$ and/or $\sigma^2 \to 0$. To achieve this purpose, we choose the following $\mathbf{Q}^{-1/2}$ matrix \cite{Gohain2022EBICR}
\begin{equation}
    \mathbf{Q}^{-1/2} = \begin{bmatrix}
    \sqrt{\frac{L_B}{N}}\sqrt{\frac{\hat{\sigma}_\mathcal{I}^2}{\hat{\sigma}_0^2}}\mathbf{I}_{k_BL_BL} & \mathbf{0}\\
    \mathbf{0} & \sqrt{\frac{L_B}{N}}\left(\frac{\hat{\sigma}_\mathcal{I}^2}{\hat{\sigma}_0^2}\right)
    \end{bmatrix},\label{eq:Q_EBIC_R}
\end{equation}
where $\hat{\sigma}^2_0 = \lVert \mathbf{y} \rVert^2_2/NL$. Also for the considered generating model (\ref{eq:LR_model_2}), $\hat{\sigma}^2_0 \to$ const. as $N \to \infty$ and/or $\sigma^2 \to 0$ \cite{Schmidt_2011_Consistency_of_MDL, GOHAIN-BIC-R}.
Two important points to note here regarding the choice of the $\mathbf{Q}^{-1/2}$ matrix are: (i) The ratio $\left(\frac{\hat{\sigma}_\mathcal{I}^2}{\hat{\sigma}_0^2}\right)$ is introduced to normalize the $\hat{\mathbf{F}}_\mathcal{I}$ w.r.t. $\sigma^2$ where the factor $\hat{\sigma}^2_0$ is especially utilized to counteract the data scaling problem (as discussed elaborately in \cite{GOHAIN-BIC-R, Gohain2022EBICR}). 
(ii) The $\frac{1}{N}$ portion of the factor $\frac{L_B}{N}$ is used to normalize the FIM w.r.t. $N$. However, $L_B$ is also included as part of the normalizing term because for the mean-squared-error of $\hat{\sigma}^2$ to approach the Cram\'{e}r-Rao bound, we require that the number of measurements is much larger than the number of parameters, i.e., $NL \gg K_B L_B L$ or in other words $N/L_B \gg K_B$. Hence, we use the normalization factor $L_B/N$ instead of just $1/N$ in (\ref{eq:Q_EBIC_R}). In this way, the penalty will be a function of $N/L_B$ instead of $N$ alone (as will be seen in the subsequent steps). This novel modification helps to counteract the effects of changing $L_B$ on the performance of EBIC$_\text{R}$.

Now, using (\ref{eq:positive_definite}), (\ref{eq:FIM_matrix}), and (\ref{eq:Q_EBIC_R}) we can show that
\begin{align}
    \Big \lvert \mathbf{Q}^{-1/2} \mathbf{\hat{F}}_\mathcal{I} \mathbf{Q}^{-1/2}\Big \rvert = & \begin{vmatrix}
    {\frac{L_B}{\hat{\sigma}_0^2}}{\frac{\breve{{\mathbf{A}}}^T_\mathcal{I}{\breve{\mathbf{A}}}_\mathcal{I}}{N}} & \mathbf{0}\\
    \mathbf{0} & \frac{L_BL}{2\hat{\sigma}_0^4} 
    \end{vmatrix}\nonumber\\
    = &\ \frac{L_B^{k_BL_BL+1}L}{2(\hat{\sigma}^2_0)^{k_BL_BL+2}}\left\lvert \mathbf{I}_L\otimes{\frac{\mathbf{A}^T_{\mathcal{I}_B}\mathbf{A}_{\mathcal{I}_B}}{N}}\right\rvert\nonumber\\    
    = &\ \text{const.}\times \left\lvert\mathbf{I}_L \right\rvert^{k_B\times L_B} \left\lvert {\frac{\mathbf{A}^T_{\mathcal{I}_B}\mathbf{A}_{\mathcal{I}_B}}{N}}\right\rvert^L\nonumber\\
    = &\ \mathcal{O}(1)
    \label{eq:normalization_EBIC_R}
\end{align}
as $N$ grows large and/or $\sigma^2 \to 0$. 
%Hence, this term can be discarded without much effect on the criterion. 
Hence, this term can be removed without significantly affecting the criterion.
Next, observe that the $\ln \big \lvert \mathbf{Q}\big \rvert$ term in \ref{eq:ln_det_FIM_EBIC_R} can be expanded as follows
\begin{align}
    \ln|\mathbf{Q}| &=
    \ln \begin{vmatrix}
    \left(\frac{N}{L_B}\right)\left(\frac{\hat{\sigma}_0^2}{\hat{\sigma}^2_\mathcal{I}}\right)\mathbf{I}_{k_BL_BL} & \mathbf{0}\\
    \mathbf{0} & \left(\frac{N}{L_B}\right)\left (\frac{\hat{\sigma}^2_0}{\hat{\sigma}^2_\mathcal{I}}\right )^2
    \end{vmatrix}\nonumber\\
    &= (k_BL_BL + 1)\ln \left(\frac{N}{L_B}\right) + (k_BL_BL + 2)\ln  \left(\frac{\hat{\sigma}_0^2}{\hat{\sigma}_\mathcal{I}^2}\right ). \label{eq:ln_det_L_EBIC_R} 
\end{align}
Therefore, using (\ref{eq:normalization_EBIC_R}) and (\ref{eq:ln_det_L_EBIC_R}) we can rewrite (\ref{eq:ln_det_FIM_EBIC_R}) as
\begin{equation}\begin{split}
    \ln{\big|\hat{\mathbf{F}}_\mathcal{I}\big|}=k_BL_BL\ln\left (\frac{N}{L_B}\right) + (k_BL_BL+2)\ln  \left(\frac{\hat{\sigma}_0^2}{\hat{\sigma}_\mathcal{I}^2}\right) +\\
    \mathcal{O}(1) + \ln \left( {N}/{L_B} \right). \label{eq:ln_FIM_EBIC_R_final}
\end{split}
\end{equation}
Next, for the model prior probability term $-2\ln \Pr(\mathcal{H}_\mathcal{I})$ in (\ref{eq:MAP_estimate_final}), a similar strategy is adopted as in EBIC \cite{EBIC2008} such that $\Pr(\mathcal{H}_\mathcal{I}) \propto  {p_B \choose k_B}^{-\zeta}$, where $\zeta \geq 0$ is a tuning parameter. If $p_B$ is sufficiently large, the following approximation can be assumed $\ln  {p_B \choose k_B} \approx k_B\ln p_B$ \cite{EFIC2018}. This gives 
\begin{equation}
    -2\ln \Pr(\mathcal{H}_\mathcal{I}) = 2\zeta k_B \ln p_B + \text{const}.\label{eq:model-prior-ebic-r}
\end{equation}
Now, substituting (\ref{eq:ln_FIM_EBIC_R_final}), (\ref{eq:model-prior-ebic-r}) in  (\ref{eq:MAP_estimate_final}) and dropping the $\mathcal{O}(1)$, the $\ln \left(N/L_B\right)$ term (since independent of $k_B$), the constant and the $p(\hat{\boldsymbol{\theta}}_\mathcal{I}|\mathcal{H}_\mathcal{I})$ term we arrive at the EBIC$_\text{R}$ for BMMV
\begin{equation}
\begin{split}
    \text{EBIC}_\text{R}(\mathcal{I}) = &  NL\ln \hat{\sigma}^2_\mathcal{I} + k_BL_BL\ln \left(\frac{N}{2\pi L_B}\right)\\& + (k_BL_BL + 2)\ln  \left(\frac{\hat{\sigma}_0^2}{\hat{\sigma}_\mathcal{I}^2}\right) + 2k_B\zeta \ln p_B.
     \label{eq:EBIC_R}
\end{split}
\end{equation}
In practice, we compute the EBIC$_\text{R}$ score block-wise, i.e., EBIC$_\text{R}(\mathcal{I}_B)$ where $\mathcal{I}_B \subseteq \{1,\ldots, p_B\}$. Then the $\hat{\sigma}^2_\mathcal{I}$ can be replaced by $\hat{\sigma}^2_{\mathcal{I}_B} = \lVert \mathbf{\Pi}^\perp(\mathbf{A}_{\mathcal{I}_B}) \mathbf{Y}\rVert^2_F\big /NL$. Finally, the true block support is estimated as
\begin{equation}
    \hat{\mathcal{S}}_B = \underset{\mathcal{I}_B  }{\arg \min} \big \{ \text{EBIC}_\text{R} (\mathcal{I}_B)\big \}.
\end{equation}

\begin{algorithm}[t]
\caption{BMMV-OMP with $K$ iterations}\label{alg:B-OMP}
\begin{algorithmic}
\State \textbf{Inputs:} Design matrix $\mathbf{A}$, measurement  $\mathbf{Y}$.
\State \textbf{Initialization:} $\lVert \mathbf{a}_j\rVert_2 = 1\ \forall j$, $\mathbf{R}^0=\mathbf{Y}$, $\mathcal{S}^0_\text{B-OMP} = \emptyset$
\For{$i=1 \text{\ to\ } K$}  
	\State Next block index: $d^i = \underset{j = 1,\ldots,p_B}{\arg\max} \left\lVert\mathbf{A}[:,\mathcal{I}_j ]^T\mathbf{R}^{i-1}\right \rVert_F$
	\State Add current index: $\mathcal{S}^i_{\textrm{B-OMP}} = \mathcal{S}^{i-1}_{\text{B-OMP}} \cup \{d^i\}$
	\State Update residual: $\mathbf{R}^i = \mathbf{\Pi}^\perp(\mathbf{A}_{\mathcal{S}^i_{\textrm{B-OMP}}})\mathbf{Y}$
\EndFor
\State \textbf{Output:} B-OMP generated block index sequence $\mathcal{S}^K_\text{B-OMP}$
\end{algorithmic}
\end{algorithm}

\begin{algorithm}[b]
\caption{Model selection combining EBIC$_\text{R}$ with B-OMP}\label{alg:OMP_ebicR}
\begin{algorithmic}
\State Run B-OMP for $K$ iterations to obtain  $\mathcal{S}^K_\text{B-OMP}$
\For{$k_B=1 \text{\ to\ } K$}  
\State $\mathcal{I}_B = \mathcal{S}^{k_B}_\text{B-OMP}$
\State Compute EBIC$_\text{R}(\mathcal{I}_B)$
\EndFor
\State Estimated true block support: $\hat{\mathcal{S}}_B = \underset{\mathcal{I}_B}{\arg\min} \{$EBIC$_\text{R}(\mathcal{I}_B)\}$
\end{algorithmic}
\end{algorithm}

%---------------------------%
\section{Simulation Results}
%---------------------------%
In this section, we provide numerical simulations to highlight the performance of EBIC$_\text{R}$ for model selection in BMMV models. We consider the linear model $\mathbf{Y} = \mathbf{A}\mathbf{X} + \mathbf{W}$, where the design matrix $\mathbf{A}$ is generated with independent entries following normal distribution $\mathcal{N}(0,1)$. The cardinality of the true block-support $\mathcal{S}_B$ is chosen to be $K_B = 4$. Also, without loss of generality, we assume $\mathcal{S}_B = [1,2,3,4]$. 
The non-zero entries in $\mathbf{X}$ are randomly assigned $\pm 1$. 
The SNR in dB = $10\log_{10}(\sigma_s^2/\sigma^2)$, where $\sigma_s^2$ and $\sigma^2$ denote signal and true noise power, respectively. The signal power is computed as $\sigma_s^2 = \lVert\mathbf{A}\mathbf{X}\rVert^2_F/NL$. 
The chosen SNR (dB) and $\sigma_s^2$ are then used to determine the noise power as $\sigma^2 = \sigma_s^2/10^{\textrm{SNR (dB)}/10} $. Using this $\sigma^2$, the elements of the noise matrix $\mathbf{W}$ are generated following $\mathbf{W}[i,j] \overset{\text{i.i.d}}{\sim}\mathcal{N}(0,\sigma^2)$. The probability of correct model selection (PCMS), i.e., $\Pr(\hat{\mathcal{S}}_B=\mathcal{S}_B)$ is evaluated over $1000$ Monte Carlo trials. 
At each Monte Carlo trial, a new design matrix $\mathbf{A}$ is generated in order to preserve the randomness in the data. 
For predictor/subset selection, BMMV-OMP (B-OMP) 
\cite{BMMV-OMP, kallummil2022GRRT} (Algorithm \ref{alg:B-OMP}) is utilized because of its ease of use and broad application.
%To maintain randomness in the data, a new design matrix $\mathbf{A}$ is generated at each Monte Carlo trial. BMMV-OMP (B-OMP) \cite{BMMV-OMP,kallummil2022GRRT} (Algorithm \ref{alg:B-OMP}) is used for predictor/subset selection for its simplicity and wide range of applicability. 
The performance of EBIC$_\text{R}$ is compared with GRRT and the oracle, which is B-OMP with a-\textit{priori} knowledge of the block sparsity $K_B$. Hence, the oracle provides the upper bound on the maximum achievable PCMS for any given setting. The tuning parameters chosen are $\alpha = 0.01$ for GRRT (as mentioned in \cite{kallummil2022GRRT}) and $\zeta = 1$ (EBIC$_\text{R}$) \cite{Gohain2022EBICR, gohain2022_EBICR_IEEEtrans}. 

\begin{figure}[t]
    \centering
    \includegraphics[trim={2.5cm 0cm 2cm 1cm},clip,scale=0.25]{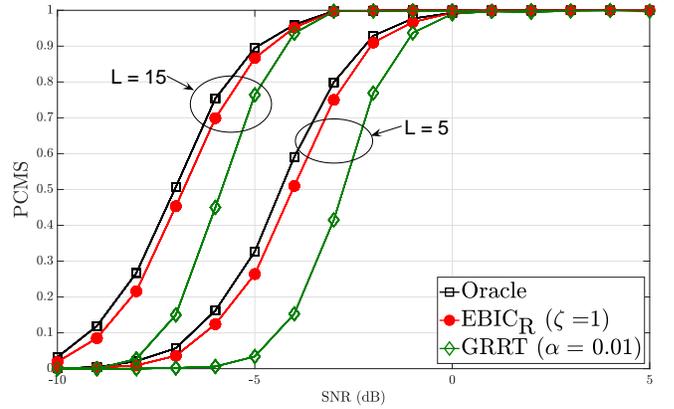}
    \caption{PCMS vs SNR (dB) for $N = 150$, $p=1000$, $L = [5, 15]$, $L_B = 10$ and $K_B = 4$.}
    \label{fig:P_vs_SNR}
\end{figure}

Fig. \ref{fig:P_vs_SNR} shows the PCMS vs SNR (dB) with $N=150$ and $p = 1000$. Since $L_B = 10$, hence, $p_B = p/L_B = 100$. Additionally, the performance is shown for two different settings of the $L$ parameter, \textit{viz.} $L=5$ and $15$ to highlight the influence of $L$ on the overall behaviour of the methods. The first clear observation is that for the considered tuning parameter setting, both EBIC$_\text{R}$ and GRRT are empirically consistent in high-SNR, i.e., PCMS $\to 1$ as SNR $\to \infty$ (or inversely $\sigma^2 \to 0$). Second, compared to GRRT, the performance curve of EBIC$_\text{R}$ is much closer to the oracle, especially for low values of SNR. Furthermore, compared to $L=5$, the oracle plot shifts toward the left when $L=15$. This indicates that increasing $L$ improves the true support recovery ability of B-OMP, which ultimately improves the model selection performance of the methods.

\begin{figure}[b]
    \centering
    \includegraphics[trim={2.5cm 0cm 2cm 1cm},clip,scale=0.25]{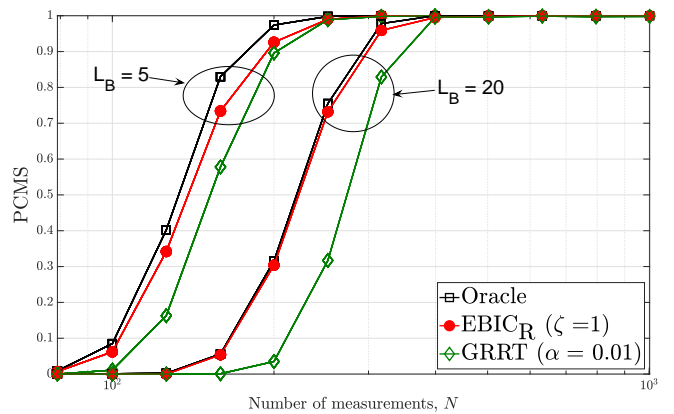}
    \caption{PCMS vs $N$ for SNR = -4 dB, $p = 5000$, $L = 5$, $L_B = [5, 20]$ and $K_B = 4$.}
    \label{fig:P_vs_N}
\end{figure}

Fig. \ref{fig:P_vs_N} presents the PCMS vs number of measurements $N$ plot. Here, a fixed value of $p=5000$ is chosen. Additionally, the performance is shown for two separate values of the $L_B$ variable, \textit{viz.} $L_B = 5$ and $20$ to highlight the impact of $L_B$ on the overall model selection performance.
A similar trend is observed here as well. Both the methods achieve empirical consistency (PCMS $\to 1$) as $N$ grows large. However,  EBIC$_\text{R}$ provides slightly better performance compared to GRRT for smaller $N$ values, and is much closer to the oracle performance. 
Furthermore, we also observe that increasing $L_B$ lowers the support recovery performance of B-OMP, which is obvious from the shift in the oracle performance towards the right. Thus, it requires more samples to achieve the same PCMS for $L_B = 20$ as compared to $L_B=5$. This ultimately lowers the overall performance of all model selection methods. 
%Hence, we can say that increasing $L_B$ affects the performance negatively.  

%-------------------%
\section{Conclusion}
%-------------------%
In this paper, we have extended the EBIC$_\text{R}$ to handle model selection in the block-sparse HD linear regression. A generalized method is developed that is applicable to all forms of the linear regression structure such as SMV, BSMV, MMV, and BMMV. The steps to arrive at the criterion are shown in detail. Simulation results show that EBIC$_\text{R}$ is an empirically consistent criterion as $N\to \infty$ and/or SNR $\to \infty$. Also. Its performance for lower SNR and $N$ values is close to the oracle behaviour. Furthermore, we also underline the manner in which the parameters $L$ and the block length $L_B$ affect the model selection performance.

% or


%\appendices

%\section{}

%\section*{Acknowledgment}


\ifCLASSOPTIONcaptionsoff
  \newpage
\fi



%\bibliographystyle{IEEEtran}
\bibliographystyle{IEEEtran}
\bibliography{refs.bib}
\end{document}